\documentclass[a4paper,10pt,twoside]{cpc-hepnp}

\usepackage{multicol}
\usepackage{graphicx}
\usepackage{booktabs}
\usepackage{amssymb,bm,mathrsfs,bbm,amscd}
\usepackage[tbtags]{amsmath}
\usepackage{lastpage}
\usepackage{CJK}
\usepackage{xcolor}

\begin{document}

\fancyhead[c]{\small Chinese Physics C~~~Vol. XX, No. X (201X)
XXXXXX} \fancyfoot[C]{\small 010201-\thepage}

\footnotetext[0]{Received 14 March 2009}

\title{The test of the electronics system for the BESIII ETOF upgrade\thanks{Supported by Chinese Academy of Sciences (No. 1G201331231172010).}}

\author{%
\quad WANG Xiao-Zhuang()$^{1,2,3}$
\quad DAI Hong-Liang()$^{2,3;1)}$\email{daihl@ihep.ac.cn}
\quad WU Zhi()$^{2,3;2)}$\email{wuz@ihep.ac.cn}
\quad HENG Yue-Kun()$^{2,3;3)}$\email{hengyk@ihep.ac.cn} \\
\quad Zhang Jie()$^{2,3}$
\quad CAO Ping()$^{1,3}$
\quad JI Xiao-Lu()$^{1,3}$
\quad Li Cheng()$^{1,3}$\\
\quad SUN Wei-Jia()$^{1,3}$
\quad WANG Si-Yu()$^{1,3}$
\quad WANG Yun()$^{1,3}$
}
\maketitle

\address{%
$^1$ Department of Modern Physics, University of Science and Technology of China, Hefei 230026, China\\
$^2$ State Key Laboratory of Particle Detection and Electronics, Beijing 100049, China\\
$^3$ Institute of High Energy Physics, Chinese Academy of Sciences, Beijing 100049, China\\
}

\begin{abstract}
It is proposed to upgrade the endcap time-of-flight (ETOF) of the Beijing Spectrometer III (BESIII) with multi-gap resistive plate chamber (MRPC), aiming at overall time resolution about 80 ps. After the entire electronics system is ready, some experiments, such as heat radiating, irradiation hardness and large-current beam tests, are carried out to certify the electronics¡¯ reliability and stability. The on-detector test of the electronics is also performed with the beam at BEPCII E3 line, the test results indicate that the electronics system fulfills its design requirements.
\end{abstract}

\begin{keyword}
BESIII,upgrade of endcap TOF,FEE,the electronics system,MRPC
\end{keyword}

\begin{pacs}
29.40.Cs
\end{pacs}

\footnotetext[0]{\hspace*{-3mm}\raisebox{0.3ex}{$\scriptstyle\copyright$}2013
Chinese Physical Society and the Institute of High Energy Physics
of the Chinese Academy of Sciences and the Institute
of Modern Physics of the Chinese Academy of Sciences and IOP Publishing Ltd}%

\begin{multicols}{2}

\section{Introduction}
The Beijing Spectrometer III (BESIII) \cite{Ref1} is a high precision general-purpose detector designed for high luminosity $e^{+}e^{-}$ collisions in the $\tau$-charm energy region at the Beijing Electron and Positron Collider II (BEPCII) \cite{Ref2}. The current BESIII ETOF detector consists of 2¡Á48 fast scintillators (BC204) readout with fine-mesh photomultiplier tubes (Hamamatsu R5924) \cite{Ref3}, the time resolution measured is 138 ps for $\pi$ (1GeV/c)). The current ETOF modules and schematics of BESIII are shown in Fig.\ref{fig1}. The secondary particles created from the multiple scatterings upon the materials between MDC endcap and ETOF lead to a high multi-hits rate (per channel), especially for electron events (~71.5\%), which lead to deteriorate time resolution \cite{Ref5}.

The proposal is approved to upgrade the current BESIII ETOF with the MRPC technology, aiming at an overall time resolution of 80 ps for MIPs. The beam tests for the MRPC prototype, together with the FEE and time digitizer (TDIG) boards, are performed at the BEPC E3 line. A time resolution of better than 50 ps can be obtained, illustrated in Ref. \cite{Ref6}\cite{Ref7}\cite{Ref8}, which verify the physical design of new ETOF.
\begin{center}
\includegraphics[width=7cm]{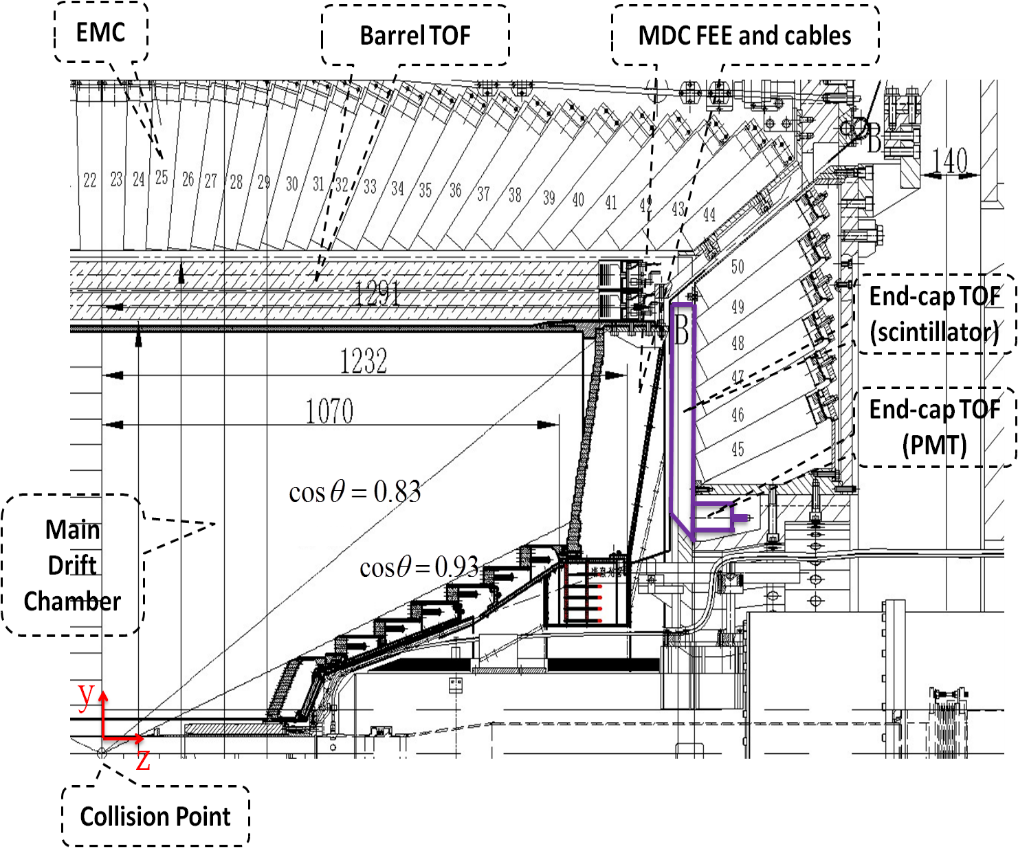}
\figcaption{\label{fig1}Schematics of BESIII ETOF}
\end{center}
In the project design, each ETOF ring has 36 overlapping MRPCs, as shown in Fig.\ref{fig2}. They are separated into 2 tiers with 18 MRPC modules in each. Each MRPC module was equiped with 12 double-end readout strips, which results in higher granularity compare to the current ETOF. The thickness of each gas box is less than 25 mm according to the space limitation. More details about the MRPC module can be found in Ref. \cite{Ref9}.
\begin{center}
\includegraphics[width=4.8cm]{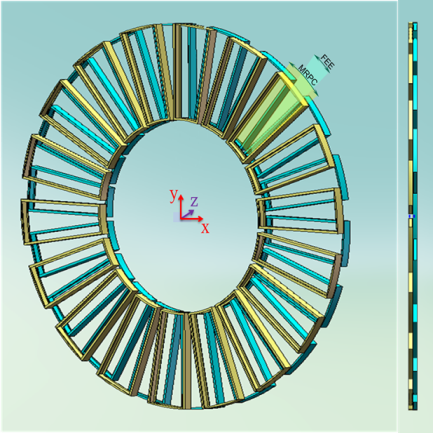}
\figcaption{\label{fig2}Structure of BESIII ETOF ring}
\end{center}
In this paper, we report the tests of the electronics system in which The Front End Electronics (FEE) , CLOCK, Fast control board and calibration-trigger-threshold-power (CTTP) modules are new-designed and firstly incorporated. In the following sections, the readout electronics and data acquisition system are introduced. Then the tests of the electronics including parameter adjustment, the heat radiating, irradiation hardness, and large-current beam are presented. The on-detector test of the electronics and its results are described and discussed in details. A conclusion is given at the end.
\section{Readout electronics and data acquisition system}
The readout electronics is mainly composed of the FEEs, 9U VME and NIM crate, together with the data acquisition system. The FEE makes use of the NINO chip developed by the ALICE-TOF group \cite{Ref10}. Each FEE module features 24 differential input channels and outputs correspondent LVDS signal with the signal charge encoded in its width. The timing accuracy RMS can be better than 15 ps for each channel when the input charge is larger than 100 fC \cite{Ref11}. The FEE board is fixed on the surface of the aluminum gas box which contains the MRPC module in order to reduce the input capacitance. A flexible printed circuit is designed to connect the MRPC module output with proper impedance (54 ohm). The connector (QSS-025-01-L-D-A-K and QSS-025-02-L-D-RA-MTI) with 86 pins and the shielded differential cable are used to connect the FEE and the TDIG \cite{Ref12}\cite{Ref13}, which are aiming at reducing the time jitter from signal transmission and ensuring the signal quality. In order to suppress the noise, the FEE are coated with aluminum shielding boxes. The picture of FEE and shield boxes are shown in Fig.\ref{fig3}.
\end{multicols}
\ruleup
\begin{center}
\includegraphics[width=12cm]{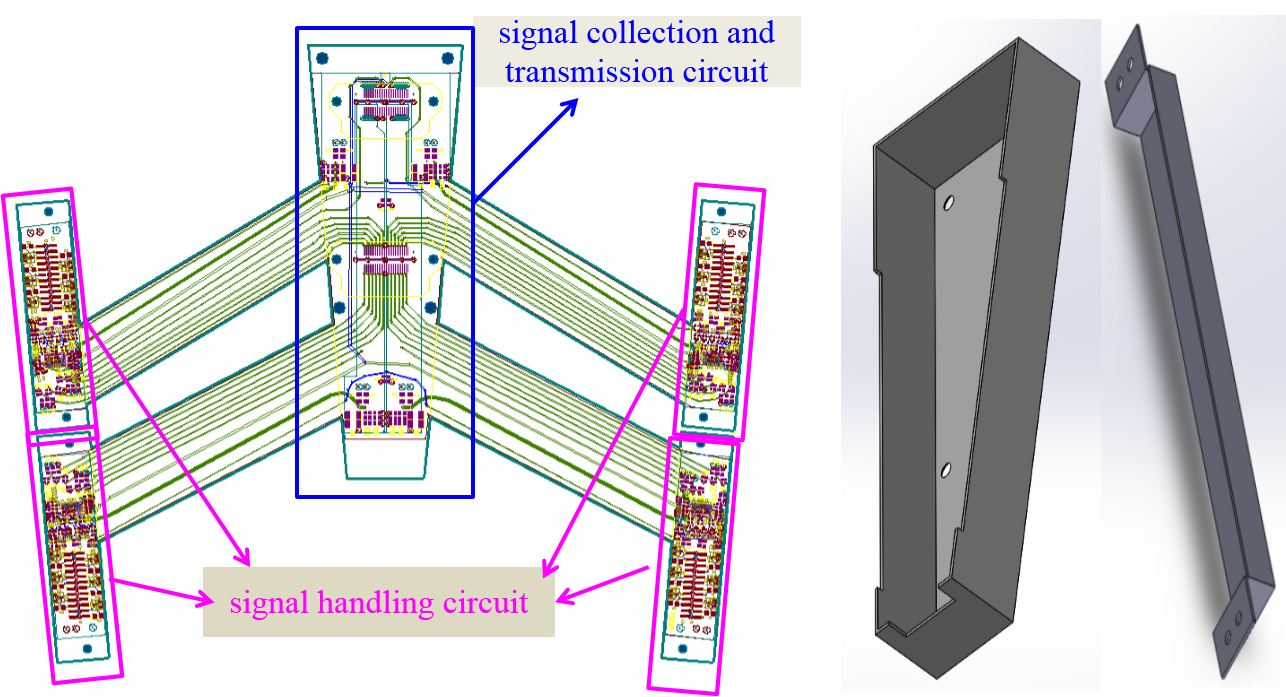}
\figcaption{\label{fig3}The schematics of FEE board and shielding boxes}
\end{center}
\ruledown
\begin{multicols}{2}
The CTTP module, which is housed in NIM crate, provides power, threshold and test signals to the FEE. It also receives the OR differential signals from the FEE and produces fast trigger signals after coincidence for charged particles identification.

The 9U VME crate consists of the VME controller of Power Printed Circiut (Power PC) board, the readout control module of ROC, the clock production module, and the TDIG module. The TDIG modules, relying on the ASIC HPTDC chip developed by the microelectronics group at CERN \cite{Ref14}\cite{Ref15}, focus on receiving and digitizing the signals from the FEE, packing the data with predefined format and uploading them to the data acquisition (DAQ) system via the VME bus. Each TDIG board integrates 72 channels with 9 HPTDC chips operating in the high resolution mode. The TDIG picture is shown in Fig.\ref{fig4}.

The data acquisition system is similar to those at the BESIII experiment, adopting the  techniques of multi-level buffering, parallel processing, high-speed VME readout and network transmission. The schematic of the readout electronics system for the ETOF upgrade is shown in Fig.\ref{fig5}.
\begin{center}
\includegraphics[width=4.5cm]{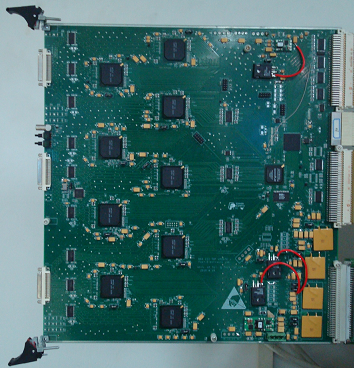}
\figcaption{\label{fig4}The picture of TDIG module}
\end{center}
There are two separately working modes for the electronics system, namely data taking mode and calibration mode, which are determined by the ROC module and the latter one is mainly used to guarantee whether the system works properly. In the mode of data taking, the ROC receives a series of control signals, such as ¡°clock¡±, ¡°trigger¡± from other system at BESIII and sends them to the module of TDIG to initialize , while in the calibration mode, the ROC generates these standard signals by itself to clarify the system. The ROC module sends signals to the CTTP to generate the test signals and to the TDIG to control the time measurements. Upon the completion for the measurements, the ROC module will produce an interrupt signal on VME bus, and then data acquisition system will read out the data from the TDIG module and deliver them to the computer.
\end{multicols}
\ruleup
\begin{center}
\includegraphics[width=12cm]{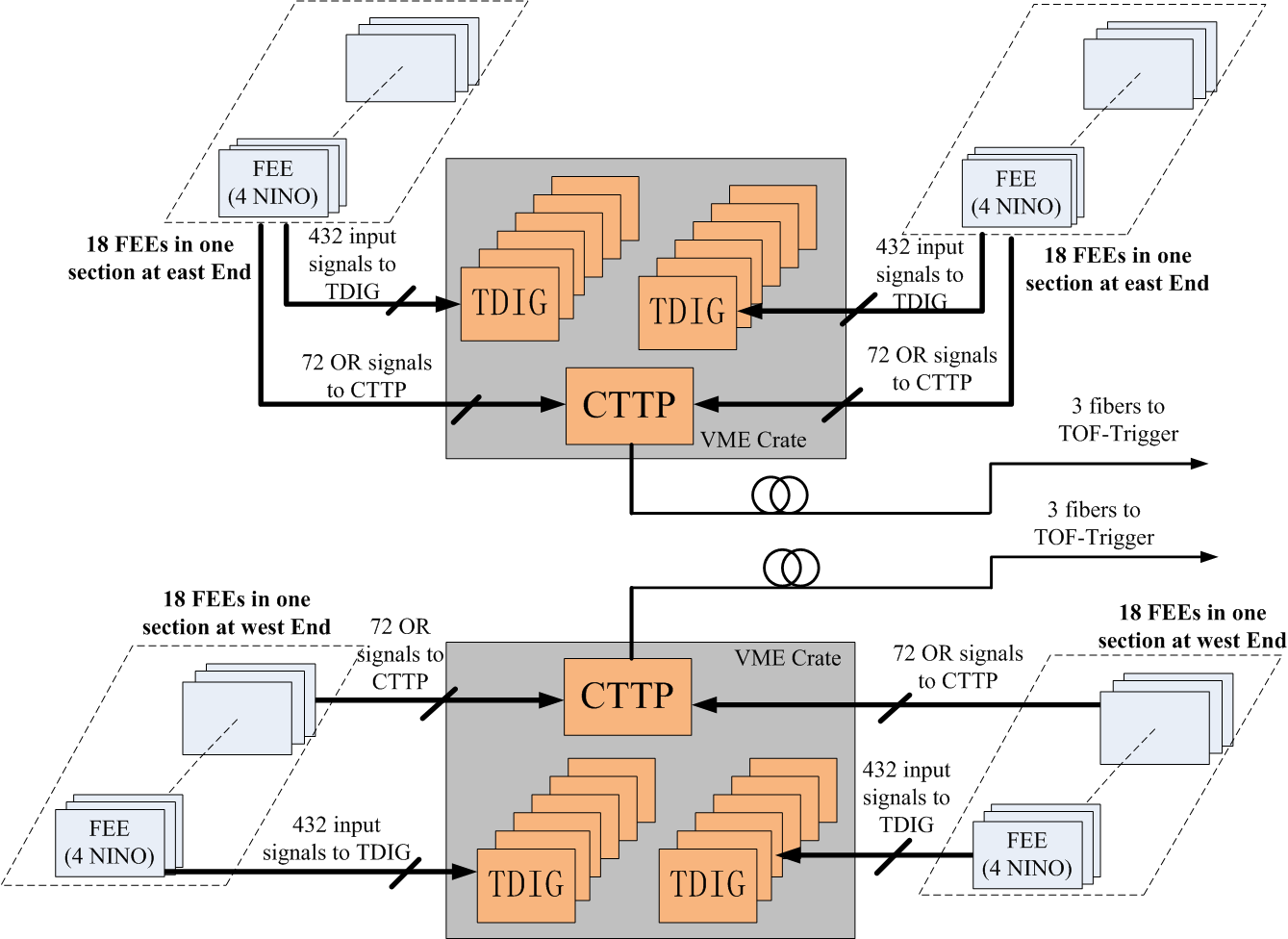}
\figcaption{\label{fig5}The schematic of readout electronics of the entire MRPC-based ETOF system}
\end{center}
\ruledown
\begin{multicols}{2}
\section{The test of electronics system}
According to the project design, Each MRPC module together with FEE will be mounted on the endcap EMC. If FEE or MRPC module is needed to repair, the endcap EMC should be firstly detached. It takes two weeks to pull out and push in the endcap EMC, so it is impossible to repair them during the data taking period. In the endcap area, the dose rates where the FEE locates are about 2000 rad after 10 years running of BEPCII. The overlapping structure of detectors isn't conducive to FEE's heat radiating. Based on the above points, the reliability tests are necessary for the electronics system, especially for the FEE.
\subsection{Parameter adjustment}
The leading and trailing time are accurately measured relying on search window and trigger latency in the TDIG module. The width of leading or trailing time distribution is about 25ns due to the clock uncertainty and the width of the signal is about 25 ns. The width of the search window is set at 1600 ns, the same as that of BESIII TOF used. After the parameter adjustment, the leading and trailing time and TOT distributions are obtained, shown in Fig.\ref{fig6}, the few hits located outside the signal region reflect the noise level.
\end{multicols}
\ruleup
\begin{center}
\includegraphics[width=8cm]{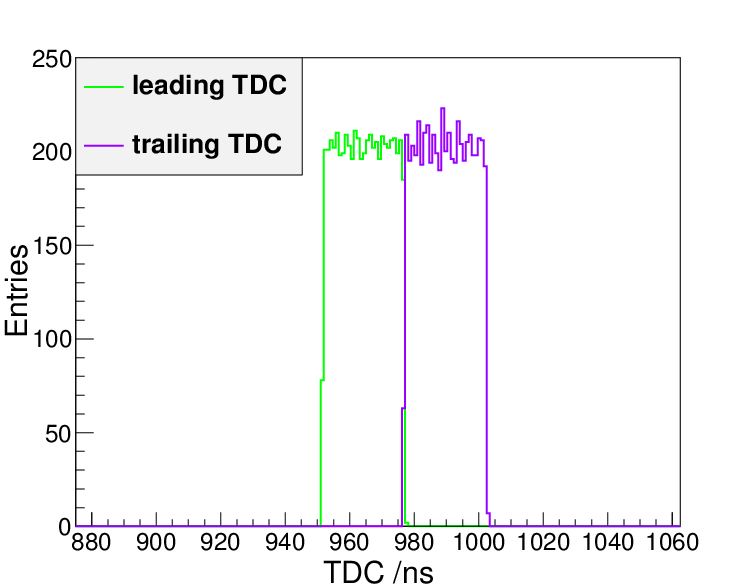}
\includegraphics[width=8cm]{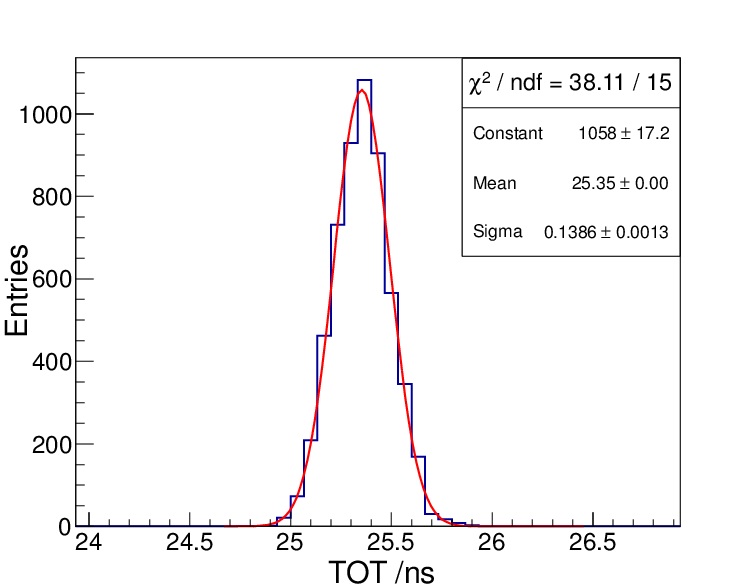}
\figcaption{\label{fig6}(left) leading and trailing time spectrum, (right) TOT distribution}
\end{center}
\ruledown
\begin{multicols}{2}
When the readout electronics and data acquisition system is ready, the first data taking is done for the calibration to check whether the electronic chain (FEE, CTTP, cables, TDIG) worked correctly. Considering the integral non-linearity correction of time measurement \cite{Ref13}, the average time resolution of electronics is about 25ps shown in Fig.\ref{fig7}, which provide the feasibility for the system-level time monitor in the future.
\begin{center}
\includegraphics[width=8cm]{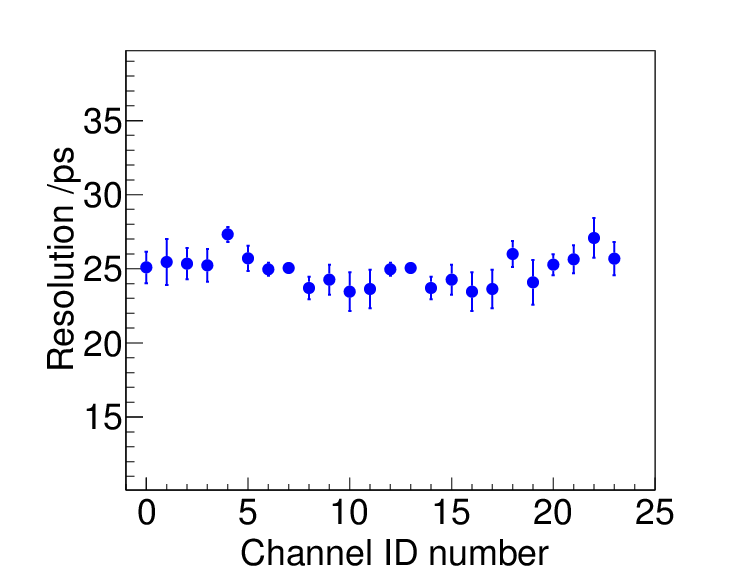}
\figcaption{\label{fig7}The time resolution versus channels}
\end{center}
\subsection{Heat radiating test}
A heat radiating test of the FEE electronincs is carried out. Two temperature sensors are used to detect the temperature during the FEE working.
The locations of two sensors is shown in Fig.\ref{fig8}: (A)near the signal collection and transmission circuit; (B)near the NINO chip in signal handling circuit. The FEE electronics combining with the MRPC module is put into a airtight box.
\begin{center}
\includegraphics[width=8cm]{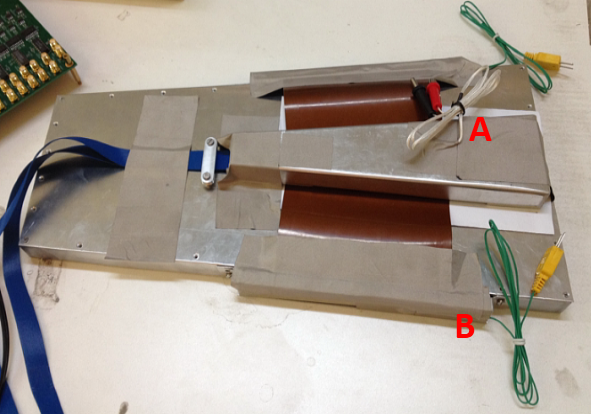}
\figcaption{\label{fig8}The picture of two sensor location in heat radiating test}
\end{center}
The test results is shown in the Fig.\ref{fig9}. At the beginning of the test, the temperature of Point B rise rapidly when FEE starts working. After the sufficient heat exchanging at certain time, the temperature of Point A and B can both keep below 30 degrees required by the BESIII experiment. On the whole, the temperature of Point A can be 1--2 degrees higher than room temperature; the temperature of Point B can be 4--5 degrees higher than room temperature. During the testing time, the current of CTTP is stable shown in Fig.\ref{fig9} and the FEE system work normaly. In the future running condition, there are dry air blowing the surface of detectors all the time, so the working of FEE should be reliable.
\begin{center}
\includegraphics[width=8cm]{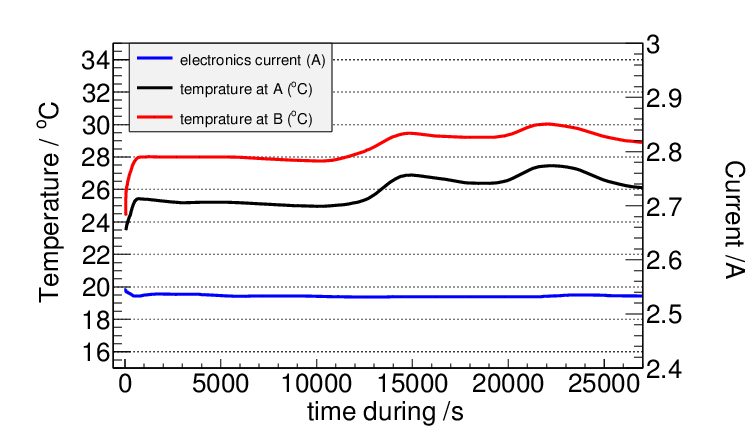}
\figcaption{\label{fig9} The results of heat radiating test}
\end{center}
\subsection{Irradiation hardness test}
Irradiation hardness test is done with ${}^{60}Co$ at Academy of Military Medical Sciences. The electronics system is exposure to radioactive source of ${}^{60}Co$. The performance of electronics is studied with different dose rates. From the results in Fig.\ref{fig10}, the resolution of electronics system keep steady with the dose increasing from 500 rad to 43500 rad.
\begin{center}
\includegraphics[width=8cm]{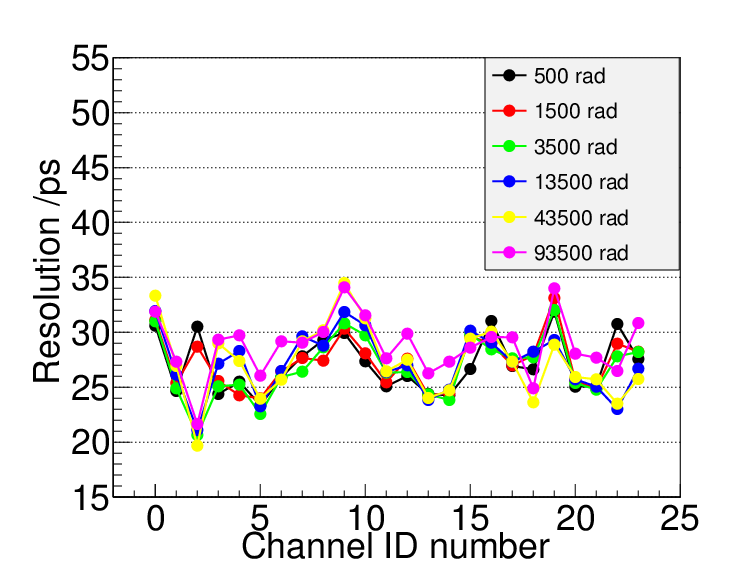}
\figcaption{\label{fig10} The resolution versus channels under irradiation}
\end{center}
\subsection{Large-current beam test}
Large-current beam test is also carried out to examine the protect circuit of FEE. The test is performed at the BEPC E2 line using 2.5 GeV incident electron beam. The FEE and MRPC module is put in the path of the beam and the high voltage of MRPC is at the normal value. The beam intensity increases from $10^{5}$ electrons to $10^{8}$ electrons at a frequency of 12.5 Hz. As the beam intensity is higher, the high voltage can not stay at normal values because the leak current of MRPC exceed the setting value. So we increase and decrease the beam intensity iteratively, keeping the hitting under normal HV condition. Finally the test lasts about two hours, the FEE can work normally, which indicate that the protect circuit of FEE is reliable.
\subsection{The on-detector test}
A on-detector test of the electronics is performed at the BEPC E3 line using the secondary particles ($e^{+/-}$, $\pi^{+/-}$, p, etc.)\cite{Ref15}. The setup of the beam test is shown in Fig.\ref{fig11}. The trigger is provided by the coincidence signal of two scintillators. The MRPC modules are placed on a movable platform and the center of different pad is moved to the trigger region.
\begin{center}
\includegraphics[width=6cm]{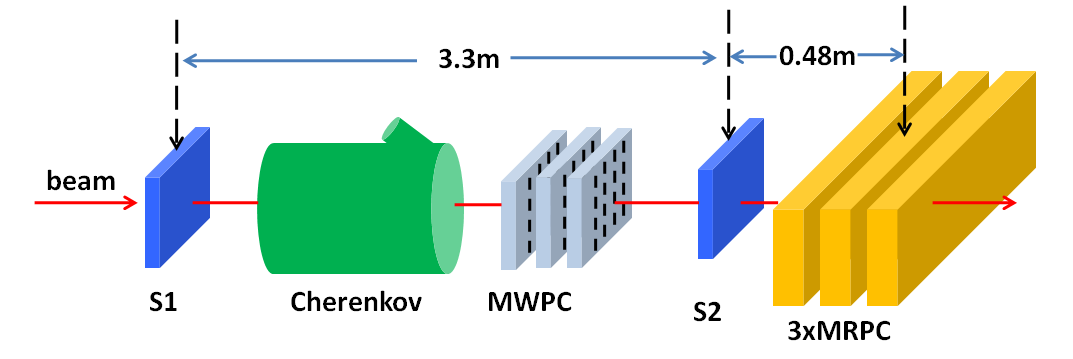}
\figcaption{\label{fig11}The setup of beam test system}
\end{center}
The high-voltages of the MRPCs are provided by N471A modules. The component of the MRPC working gas is 90\%Freon + 5\%$SF_{6}$ + 5\%iso-${C_{4}H_{10}}$ for the test. The gas flux rate is supplied at 60 ml/min to the MRPC modules. The logic diagram of the test system is shown in Fig.\ref{fig12}.
\begin{center}
\includegraphics[width=8cm]{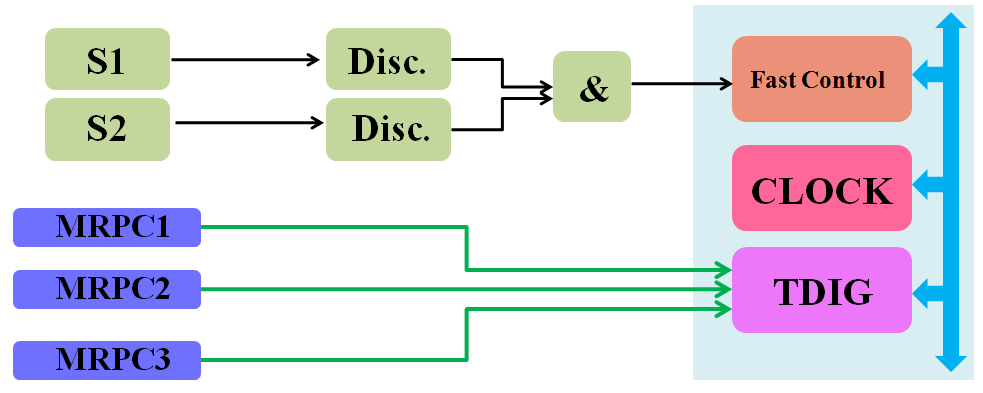}
\figcaption{\label{fig12}The module logic diagram of the test system}
\end{center}
The flight time between S1 and S2 is used to identified pion and proton, which is shown in Fig.\ref{fig13}. Since the limited statistics for pions, we slected proton events for our MRPC performance analysis.
\begin{center}
\includegraphics[width=8cm]{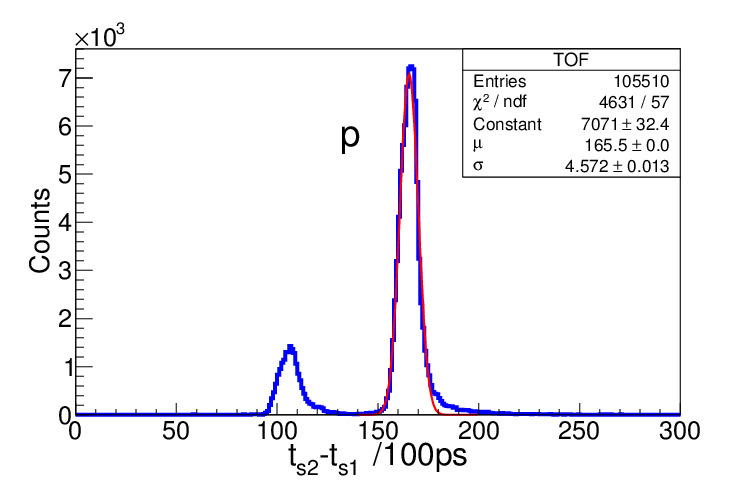}
\figcaption{\label{fig13}The particle identification of $s_{2}$ and $s_{1}$ @ 800 MeV}
\end{center}
The time measured by the MRPC is corrected for the slewing by iterations method \cite{Ref8}. The Time-TOT correlation for each MRPC module are fitted with the mean time of the other two as the reference time ($T_{r}$), then use the corrected time instead of primary time to be $T_{r}$ and draw a new Time-TOT distribution for each one which will be more exact than the primary one since they has better reference time. The time resolution of each pad is better than 50 ps, the average efficiency is almost 98\%, shown in Fig.14. The performance of detectors is consistent with previous beam tests, which indicate the electronics, including the firstly incorporated ones, meet the project requirement.
\begin{center}
\includegraphics[width=8cm]{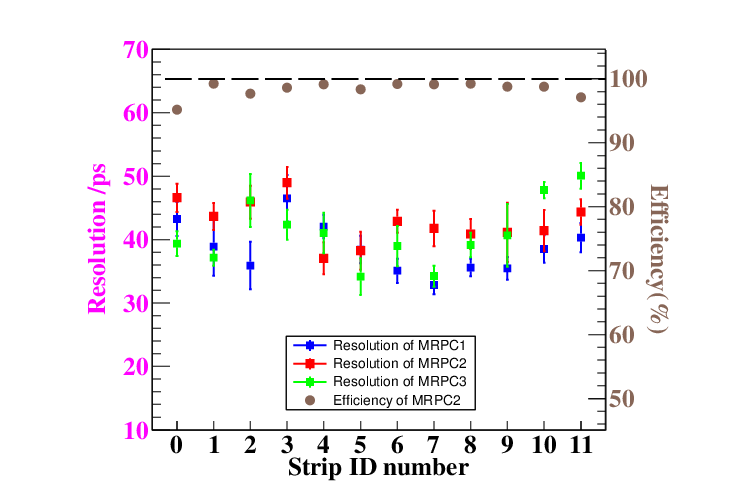}
\figcaption{\label{fig14}Time resolution and efficiency versus strip ID number}
\end{center}
\section{Conclusions}
The upgrade of the BESIII ETOF with MRPC technology has been approved. After the electronics and DAQ system is ready, the parameter adjustment has been performed firstly. Then the irradiation hardness, heat radiating and large current beam tests are carried out to prove that the electronics system, especially the FEE, is reliable and stable. The on-detector test of the electronics is also performed with beam at BEPCII E3 line, the time resolution less than 50 ps and the detecting efficiency over 98\% are obtained. These results indicate that the electronics satisfy the design requirements, and the whole system is ready for the mass production.

\end{multicols}

\clearpage

\end{document}